# Monetization as a Motivator for the Freemium Educational Platform Growth

Ilya V. Osipov [1*], Anna Y. Prasikova [1], Alex A. Volinsky [2*]

[1] i2istudy.com, Krišjāņa Barona Iela, 130 k-10, Rīga, Lv-1012, Latvija

[2] Department of Mechanical Engineering, University of South Florida, 4202 E. Fowler Ave., ENB118, Tampa FL 33620, USA

[*] Corresponding authors. Email: volinsky@usf.edu; Phone: (813) 974-5658; Fax: (813) 974-3539 (Alex A. Volinsky); Email: ilya@i2istudy.com (Ilya V. Osipov)

## Abstract

The paper describes user behavior as a result of introducing monetization in the freemium educational online platform. Monetization resulted in alternative system growth mechanisms, causing viral increase in the number of users. Given different options, users choose the most advantageous and simple ones for them. System metrics in terms of the K-factor was utilized as an indicator of the system user base growth. The weekly K-factor almost doubled as a result of monetization introduction. Monetization and viral growth can be both competing and complementary mechanisms for the system growth.

## Introduction

Online collaborative platform for learning foreign languages and improving spoken language skills has been developed. Basic idea is that in spite of the learned grammar, students are lacking live human interactions with foreigners to improve their oral communication skills (Gress, 2010). Demands of the students learning Spanish were investigated. It was found that live communication with native speakers significantly improved their spoken language skills. These native speakers don't necessarily have to be professional language teachers if provided with simple teaching materials. (Reichelta, 2014; Jin, 2011; Hadwin, 2006; Chen, 2014) Spanish-speaking students are also eager to teach Spanish in return for learning English. The authors used the idea of time banking (Válek, 2013; Marks, 2012) to track how much time is spent teaching and learning foreign languages in the system. The developed platform uses direct audio-video live communication between the users, along with the step-by-step teaching instructions combined in the online computer program with time tracking (Osipov, 2013). The program separately tracks teaching and learning time usage and provides the means to select and connect users. This online collaborative platform for learning and teaching foreign languages is called i2istudy, and is based on knowledge sharing in a social network (Hsu et al., 2007). Audio-video live communication between the users allows sharing emotions (Janssen et al., 2014), resulting in improved learning outcomes.

## Freemium system description

The i2istudy system, which grew to over 40,000 users, was used to conduct reported experiments and measure corresponding parameters. In this paper the user behavior was affected by changing online software system parameters to maximize the number of the system users (Eyal, 2014). Despite the fact that the platform was envisioned as a free application, some monetization elements were stipulated (Skok, 2009). These paid features are used by a small percentage of users needed to maintain the system. This free service model with some paid features is called “freemium”, and is a popular economic model in computer and mobile games, software-as-a-service solutions and web applications. Typical functions of a freemium product aimed at the mass market are (Fields, 2012):

1. Basic main function of the application, the “core” cycle, for which the users decide to use the application (in this case it is learning foreign languages with native speakers).

2. Monetization function as an additional feature, which involves the most excited users. In the freemium products this feature is obviously not for all involved users.

3. Retention is the cycle of users leaving the system with subsequent return. For their successful return and retention different actions are utilized, ranging from notifications by e-mail, messages in social networks and other communication channels with reminders about the ongoing events in their absence, along with gamification elements (Pe-Than, 2014, Seaborn and Fels, 2015) as additional motivating factors (Serrano-Cámara, 2004).

4. Viral function of the existing system users inviting new users from the external environment (e-mail, social networks, blogs, forums, personal web sites, applications and other communication channels), including new user accommodation.

The authors have studied the statistics of the user behavior in the developed shareware pseudo-free system. While paid features of a typical freemium application are used by 3-10% of users, statistics collected in the i2istudy system was different. From the control standpoint, freemium application is an open system consisting of users and centralized software, which controls human behavior in the system. Herewith, the users are free to participate in the system mechanics or refuse to participate anytime. In our case the purpose of control and optimization is to maximize the impact of several parameters, which characterize the system. These parameters are the growth of the system user base, reflected by the K-factor [10], average user time spent in the system over a certain period of time (retention) and system monetization (percentage of paying users and how much money they spend).

It is clear that trying to maximize several parameters simultaneously is not a correct approach. Thus, the first objective is to grow the system in terms of the daily or weekly active user and the virality in terms of the K-factor, which reflects how existing users invite new users. The second objective is the system monetization at later development stages when there are enough users. However, it was decided to implement monetization during the system initial growth to study if the existing audience is ready to pay for the premium features.

Time in minutes available to learn foreign languages from native speakers is the system internal currency. Figure 1 shows the diagram of the system before monetization, where users could only earn minutes in the system by inviting friends and teaching other users. The system consists of the user interface and the users. A new feature was added to positively affect user behavior. The hypothesis was that some users would utilize the new feature. If a large percentage of users utilize the new feature by willingly purchasing minutes, the system is monetized quite well. If this percentage is low, the experiment was unsuccessful.

**Monetization**

The control mechanism was realized by changing the interface and some programming algorithms to motivate users by a new function. As a result, the system was changed from the state I in Figure 1 to the state II in Figure 2, which includes additional branch of the possible user behavior algorithm to purchase minutes. As mentioned by Seufert (2013), factors, such as virality, gamification and monetization can enhance each other, increasing user involvement, or be antagonistic and compete for user attention.  Thus, different behavioral scenarios, labeled A, B, C and D in Figure 2 can either help or interfere with each other, depending on the situation.

In this case, the system interface was utilized to attract user attention to scenarios A, B and C to improve user involvement in the process. Besides, scenarios B and C represent the core cycle of the application, where users spend most of the time teaching and learning from each other. Scenario A is the part of the interface where users can invite friends to expand the user base and receive bonuses (gamification elements to motivate users to certain actions). Curiously enough, monetization (scenario D) can drive virality and retention, since users can acquire premium features (also available for money) by participating in viral programs to invite friends. The price of these features reflected in real valuable dollars is encouraging users by clearly demonstrating their value. Users motivated by the value of the new function reflected in dollars can be easily encouraged to earn it by performing certain actions of inviting friends, for example.

To test this hypothesis the authors added monetization during the second month after the system launch. Users could now purchase minutes in the system using real money in addition to earning minutes by teaching or inviting friends. Figure 3 shows the user interface option of purchasing minutes in the system using real money. This new feature was supposed to motivate users to earn more minutes in the system by teaching their native language and inviting friends. These are two available options A and C in Figure 2 to earn system minutes without spending real money. On the other hand, the new feature should have caused real sales, allowing the developers to earn money (Park, 2011). As mentioned earlier, statistics of successful freemium application sales shows that 3-10% of the users purchase premium features (Ellis, 2014).

Attracting new users was marketed in the Facebook social network by placing ads in all four languages supported by the system (English, German, Spanish and Russian). As a result of these ads: “Teach your native language and learn English, German, Spanish or Russian in return for free”, 40,000 users registered in the system. Over 1,000-1,500 users visited the site daily. Table 1 lists the main countries, from which users have registered in the system. Users from 145 countries have used the system, signifying the demand for online language education, especially in developing countries. However, since the real money feature implementation, only two $10 sales took place. This is regardless of the fact that about 24,000 users were registered in the system at the time, and 36% of them were involved in the process of spending and earning virtual system currency, minutes (not all user activity in the system results in spending or earning minutes). Thus, only two users were involved in monetization, which is by no means satisfactory for a freemium product. Please keep in mind that the option to purchase minutes was actively marketed. In particular, special prompts were added to the system interface, along with informing all users of this new feature by email (Cohen, 2014).

However, regardless of the lack of the virtual currency sales, the demand for other ways to earn minutes increased significantly. In particular, users started to send invitations to their friends more actively. They were offered 30 minutes of bonus system time for each invited registered user. User involvement in the teaching process also increased. Additional 8% of the users indicated that they are willing to teach their native language. Before this profile parameter was filled by 47% of the users. This monetization approach resulted in 55% of the users involved in the teaching process. Thus, adding the option to purchase minutes indirectly improved system parameters of virality and retention, but had no effect on monetization.

**System metrics**

While there are different ways to assess user involvement and immersion (Jennett et al., 2008), the authors were particularly interested in how existing users invite their friends, which was assessed by calculating the K-factor. The K-factor is the ratio of the virally attracted new users and active existing users over a certain time. Kim (2000) identified five types of users: visitor, novice, regular, leader and elder. To calculate the K-factor, one could use only the new users ($NU$), all users during a certain period ($U$), or only active users ($AU$). The authors used only active users to get more accurate results, accounting for the users who passed the “membership ritual” (Kim, 2000). Some sources use only novice users as the base, comparing all virally attracted users with all types of users (Kim, 2000; Rigatuso, 2014). The authors did not consider this approach, since all active users contribute to virality, not just the novice users. The weekly K-factor was calculated as:

$$K_{factor}^{weekly} = \frac{wIU}{wAU} \cdot 100\% \quad (1),$$

where $wIU$ is the number of invited users and $wAU$ is the number of active users in a given week. The term K-factor has been taken from epidemiology, where a virus having a K-factor of 1 is in a “steady” state of neither growth nor decline, while a K-factor greater than 1 indicates exponential growth and a K-factor less than 1 indicates exponential decline.

The K-factor, commonly known in the literature as the viral coefficient (Fong, 2014), can also be calculated as the number of invitations sent by each user multiplied by the conversion percentage of each invitation into a new user (Cohen, 2014). For example, if the average number of invitations is 5, and the conversion factor into new users is 20%, then the K-factor = 5·0.2 = 1 (Seufert, 2013). Figure 4 shows the weekly K-factor dynamics with time. To make the K-factor calculations more objective the raw data was corrected to exclude invitations sent by the system developers (Table 2).

The K-factor regardless of the time period over the whole system lifetime is often called the global K-factor, which is calculated as follows. The conversion percentage is calculates as the number of users, who accepted the invitation, $IU$, divided by the number of sent invitations, $I$:

$$IPi = \frac{IU}{I} \cdot 100\% \quad (2)$$

The average number of invitations sent by each user, $AiPU$, is the ratio of the total number of invitations, $I$, and the total number of users, $U$:

$$AiPU = \frac{I}{U} \quad (3)$$

Then the K-factor is calculated as:

$$K_{factor} = AiPU \cdot IPi \quad (4)$$

To assess the system growth the authors used the weekly K-factor (Reichheld, 2003). Mass mailing informing users of the opportunity to purchase minutes in the system started on July 28th, 2014. This action caused a spike in the number of sent invitation to get extra 30 minutes in the system for each invited registered user, which affected the weekly K-factor in Table 2 and Figure 4. As seen in Figure 4, the weekly K-factor increased from 2.01±0.84% to 3.89±0.73% after the new feature announcement. The K-factor growth continued until the system was closed on September $1^{st}$ 2014 for renovation.

To test the hypothesis whether the K-factor changed as a result of monetization, statistical analysis was utilized. The null hypothesis was that the weekly K-factor did not change. The K-factor data before monetization were used as the expected values. The actual K-factor data for comparison were taken after monetization implementation. The calculated p-value was 0.01%, thus the hypothesis that the K-factor did not change as a result of monetization was rejected.

**User feedback and system development**

It is understandable that the offered monetization model did not work, although it positively affected the system overall. The authors decided to leave this monetization available, but only as the motivating factor for users to earn virtual currency by other means. For real monetization the authors plan to utilize other methods. For example, in addition to the virtual system currency in minutes, there will be user accounts with real money, along with the option to learn and teach for real money, based on the user preferences.

In the event that users will choose the option to teach and learn for money, real money will be credited to the teacher account and debited from the student account. The system will take a commission in this case. The opportunity to teach for money will only become available to teachers after a certain number of teaching hours for virtual system “minutes”. The teacher will also have to have a good ranking generated from the student assessments at the end of each lesson.

These concussions were reached after asking the system users. The site contains the red sticky button “Give us your feedback” on the bottom of the screen (Figure 3), which is actively utilized by the users. In 6 months the users left 1824 messages in 4 languages. This feedback was analyzed by dividing it into the following groups (Seufert, 2013; Mäntymäki, 2011):

36% contained praise and thank you messages, such as: “Thank you, this is cool”;

24% asked about how the system works, or had specific questions about the interface;

18% related to technical issues with configuring the hardware (the microphone and the camera) and/or inability to establish audio-video connection with other users;

11% contained negative feedback in terms of the complaints connected with other users’ behavior, or requests to close their account;

8% contained ideas related to teaching only for money and unwillingness or inability to participate in the language teaching time exchange;

4% specific suggestions for system development typically associated with adding other languages: Arabic, French, Italian and Ukrainian;

35% other unclassified feedback.

Analysis of this feedback and personal communication with the system users lead to a hypothesis that the majority of users prefer to pay for learning languages due to various reasons. They did not want to participate in the time exchange, which looked so appealing to the system developers. Thus, it was decided to develop real monetization, reflecting natural organic user needs. There are 8% of users who prefer to pay for their studies due to the unwillingness to teach or lack of demand for their native language. These 8% also include users who want to teach their native language for profit, even small. As a result there was a decision made to add the second currency into the system in terms of the real money and allow users to tech for profit. The system will allow adding real money to the system, and getting paid as well.

**Conclusions**

In conclusion, the initial attempt to monetize the system by changing the user interface and adding a new feature affected the system in an unexpected way. Adding the option to purchase internal system currency (minutes) using real money did not result in significant sales, but motivated users to utilize alternative ways to earn this virtual currency. It was expected that the users would either utilize or not utilize this option without changing the system performance parameters. However, this resulted in the unexpected growth in viral mechanics of inviting friends. The rate of attracting new users increased as a result of the offer to purchase virtual currency. Thus it was observed that direct motivation (Hung et al., 2011) for the certain action within the system was ignored, but changed other indirect parameters (system growth in this case). It is quite clear and obvious that given alternative ways to earn virtual currency, users choose the most advantageous and simple ones for them. Viral system parameters improvement just due to the unrealized option to purchase minutes is an interesting effect, which is worth studying.

**Figure Captions**

Figure 1. Diagram of the system state I before monetization, where users can only earn minutes by inviting friends and teaching other users.

Figure 2. Diagram of the system state II after monetization, where users can purchase minutes using real money.

Figure 3. System interface with the option of purchasing minutes using real money.

Figure 4. The weekly K-factor affected by the new monetization feature introduced on 6/28/2014.

**Table Captions**

Table 1. The list of countries from which the majority of the users registered.

Table 2. Data used for the weekly K-factor calculations in Figure 4.

**Figures**

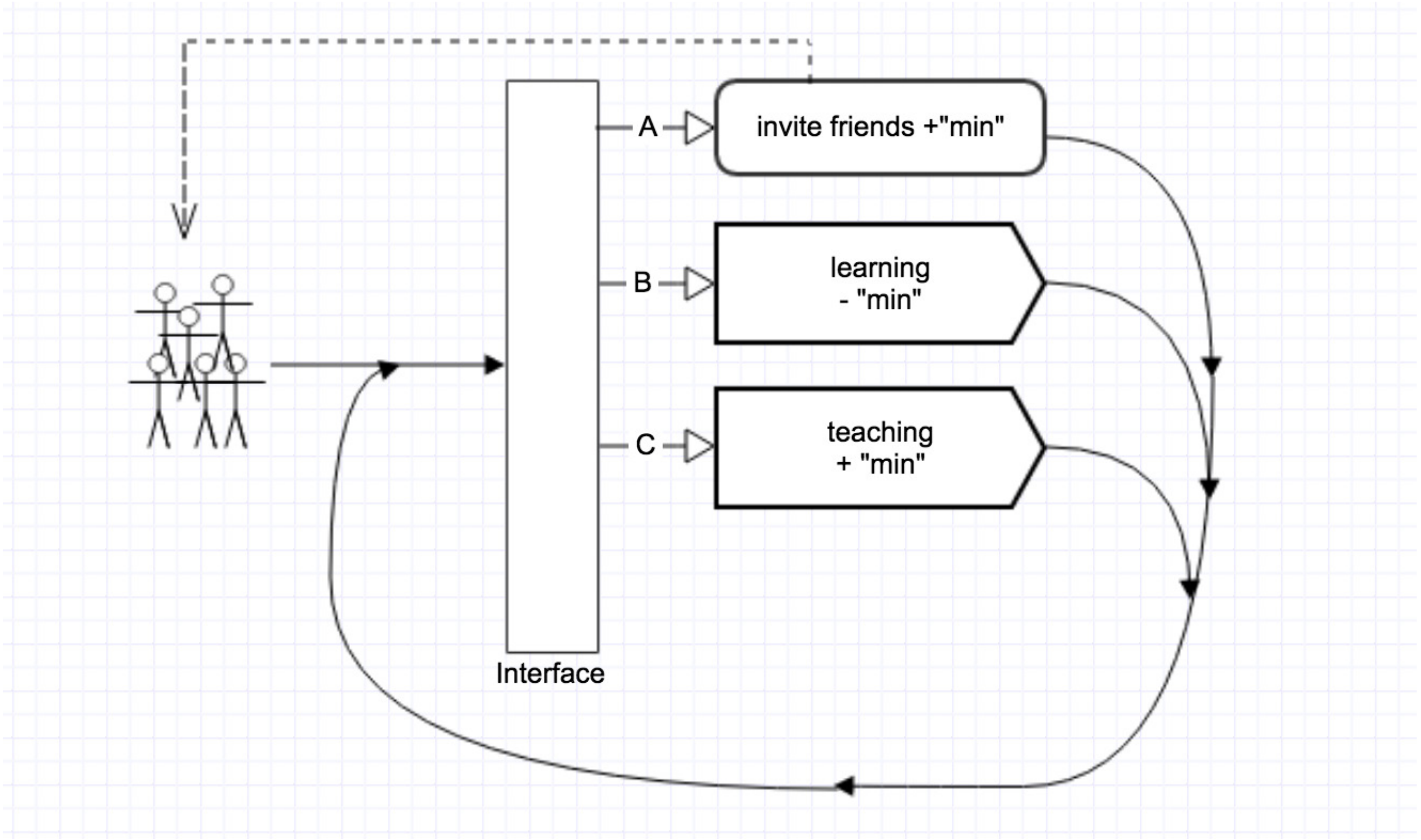

Figure 1. Diagram of the system state I before monetization, where users can only earn minutes by inviting friends and teaching other users.

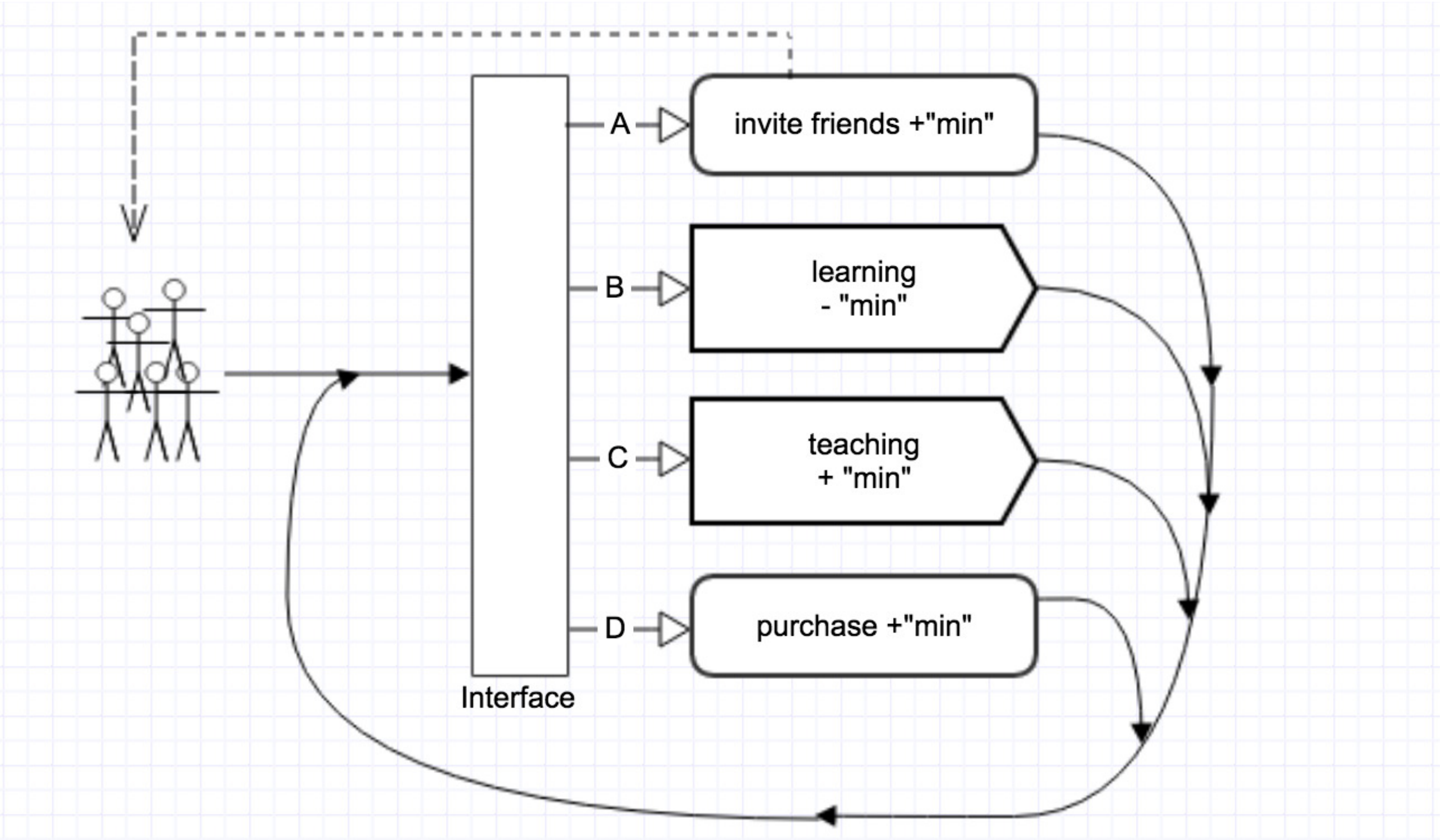

Figure 2. Diagram of the system state II after monetization, where users can purchase minutes using real money.

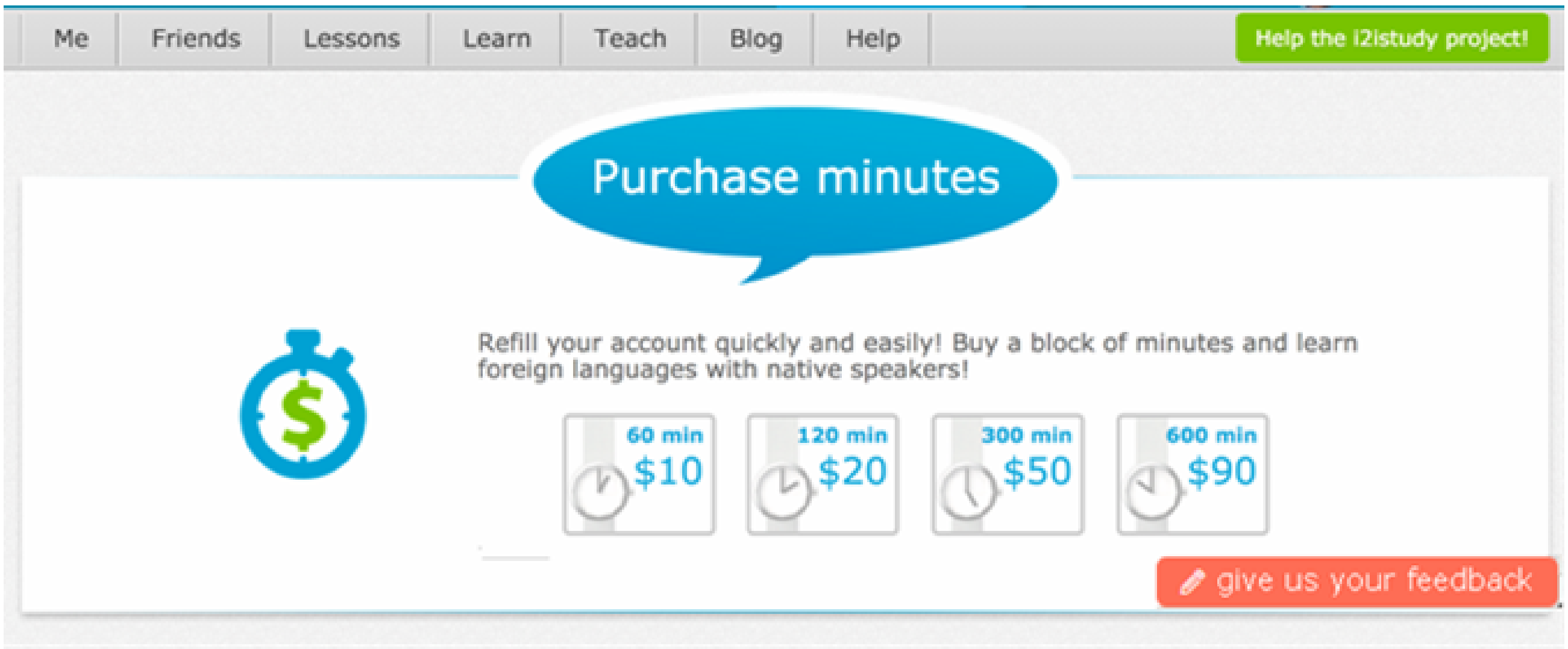

Figure 3. System interface with the option of purchasing minutes using real money.

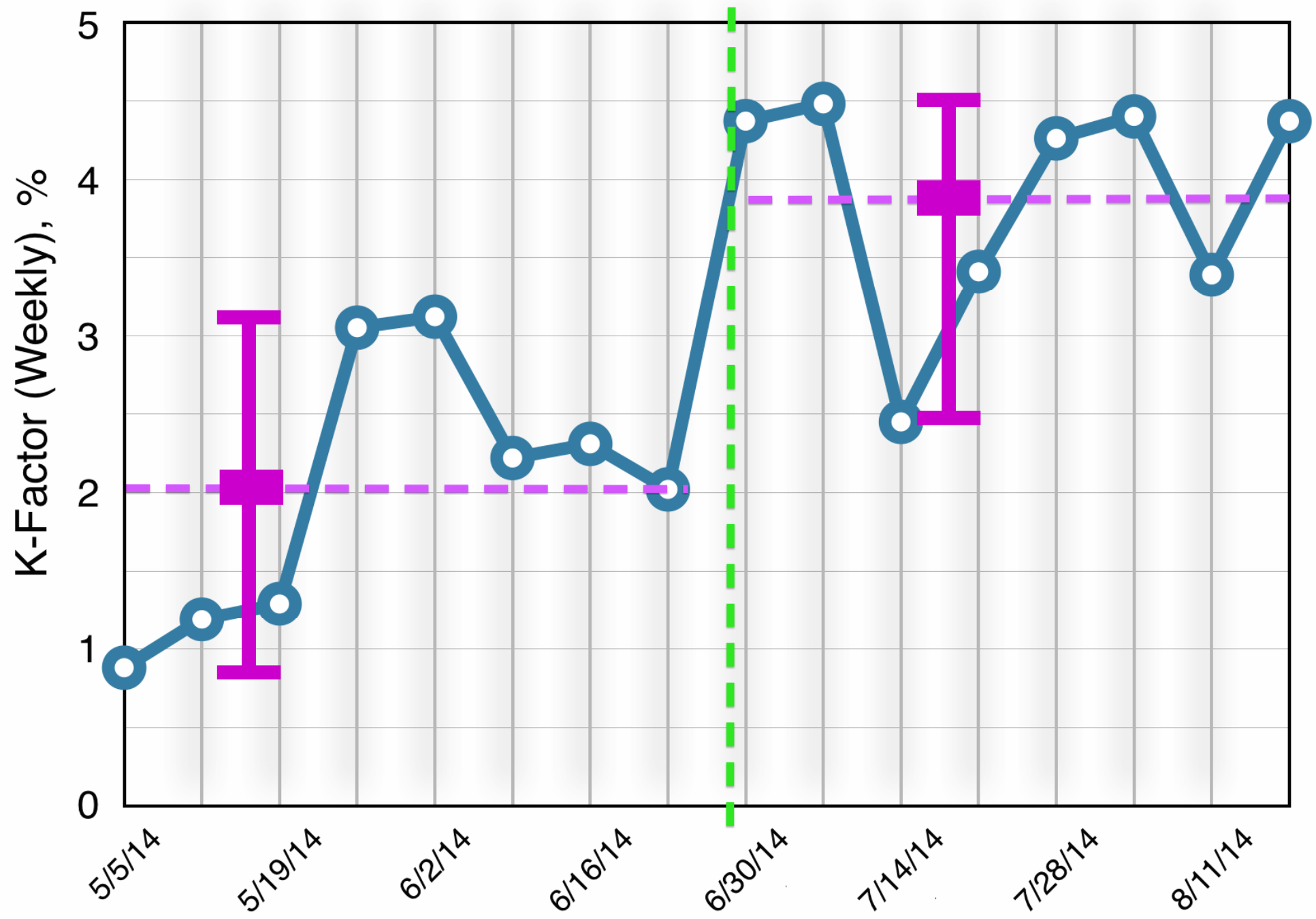

Figure 4. The weekly K-factor affected by the new monetization feature introduced on 6/28/2014.

Table 1. The list of countries from which the majority of the users registered.

| Abbreviation | Country | Number of users |
|---|---|---|
| UA | Ukraine | 7754 |
| RU | Russian Federation | 6204 |
| PK | Pakistan | 5715 |
| KZ | Kazakhstan | 3255 |
| VE | Venezuela, Bolivarian Republic of | 3124 |
| PH | Philippines | 2075 |
| BY | Belarus | 2035 |
| EG | Egypt | 1526 |
| MX | Mexico | 581 |
| MD | Moldova, Republic of | 479 |
| US | United States | 466 |
| TR | Turkey | 398 |
| MA | Morocco | 395 |
| DE | Germany | 338 |
| CZ | Czech Republic | 255 |
| FR | France | 231 |
| GB | United Kingdom | 215 |
| CO | Colombia | 192 |
| AM | Armenia | 175 |
| KG | Kyrgyzstan | 161 |
| ES | Spain | 158 |
| IN | India | 150 |
| GE | Georgia | 149 |
| IT | Italy | 148 |
| SA | Saudi Arabia | 143 |
| DZ | Algeria | 138 |
| AZ | Azerbaijan | 134 |
| AE | United Arab Emirates | 130 |
| BO | Bolivia, Plurinational State of | 122 |
| RO | Romania | 118 |
| EE | Estonia | 104 |
| IL | Israel | 88 |
| UZ | Uzbekistan | 88 |
| PL | Poland | 88 |
| LV | Latvia | 81 |
| TJ | Tajikistan | 79 |
| SK | Slovakia | 57 |
| BE | Belgium | 56 |
| CN | China | 55 |
| HU | Hungary | 52 |
| GR | Greece | 52 |
| | Other | 1286 |

Table 2. Data used for the weekly K-factor calculations in Figure 4.

| Week | 21.04 - 27.04 | 28.04 - 04.05 | 05.05 - 11.05 | 12.05 - 18.05 | 19.05 - 25.05 | 26.05 - 01.06 | 02.06 - 08.06 | 09.06 - 15.06 | 16.06 - 22.06 | 23.06 - 29.06 | 30.06 - 06.07 | 07.07 - 13.07 | 14.07 - 20.07 | 21.07 - 27.07 | 28.07 - 03.08 | 04.08 - 10.08 | 11.08 - 17.08 | 18.08 - 24.08 |
|---|---|---|---|---|---|---|---|---|---|---|---|---|---|---|---|---|---|---|
| Weekly Active Users (wAU) | | | 297 | 801 | 867 | 979 | 1080 | 1213 | 1827 | 2126 | 1763 | 2624 | 2572 | 1924 | 1716 | 1810 | 1576 | 947 |
| New Active Users (wNU) | | | 239 | 575 | 572 | 643 | 695 | 776 | 1358 | 1464 | 1140 | 1881 | 1643 | 1129 | 998 | 1066 | 824 | 358 |
| Invited Active Users (wIU) | | | 5 | 9 | 10 | 32 | 54 | 81 | 85 | 41 | 73 | 68 | 52 | 65 | 77 | 78 | 52 | 42 |
| K-Factor = wIU/wAU, % | | | 0.88 | 1.19 | 1.29 | 3.05 | 3.12 | 2.22 | 2.31 | 2.02 | 4.37 | 4.48 | 2.45 | 3.41 | 4.26 | 4.4 | 3.39 | 4.37 |